\documentclass[showpacs,epsf,floats,pre]{revtex4}
\usepackage{amssymb}
\usepackage{amsfonts}
\usepackage{amsmath}

\input{tcilatex}

\begin{document}

\title{Geometrodynamics of Information on Curved Statistical Manifolds and
its Applications to Chaos}
\author{C. Cafaro}
\email{carlocafaro2000@yahoo.it}
\author{S. A. Ali}
\email{alis@alum.rpi.edu}
\affiliation{Department of Physics, State University of New York at Albany-SUNY,1400
Washington Avenue, Albany, NY 12222, USA}

\begin{abstract}
A novel information-geometrodynamical approach to chaotic dynamics (IGAC) on
curved statistical manifolds based on Entropic Dynamics (ED) is presented
and a new definition of information geometrodynamical entropy (IGE) as a
measure of chaoticity is proposed. The general classical formalism is
illustrated in a relatively simple example. It is shown that the
hyperbolicity of a non-maximally symmetric $6N$-dimensional statistical
manifold $\mathcal{M}_{s}$ underlying an ED Gaussian model describing an
arbitrary system of $3N$ degrees of freedom leads to linear
information-geometric entropy growth and to exponential divergence of the
Jacobi vector field intensity, quantum and classical features of chaos
respectively. An information-geometric analogue of the Zurek-Paz quantum
chaos criterion in the classical reversible limit is proposed. This analogy
is illustrated applying the IGAC to a set of\textbf{\ }$n$\textbf{-}%
uncoupled three-dimensional anisotropic inverted harmonic oscillators
characterized by a Ohmic distributed frequency spectrum.
\end{abstract}

\maketitle


%\section{Introduction}

\section{\textbf{Introduction}}

The lack of a unified characterization of chaos in classical and quantum
dynamics is well-known. In the Riemannian \cite{casetti} and Finslerian \cite%
{cipriani} (a Finsler metric is obtained from a Riemannian metric by
relaxing the requirement that the metric be quadratic on each tangent space)
geometrodynamical approach to chaos in classical Hamiltonian systems, an
active field of research concerns the possibility of finding a rigorous
relation among the sectional curvature, the Lyapunov exponents, and the
Kolmogorov-Sinai dynamical entropy (i. e. the sum of positive Lyapunov
exponents) \cite{kawabe}. The largest Lyapunov exponent characterizes the
degree of chaoticity of a dynamical system and, if positive, it measures the
mean instability rate of nearby trajectories averaged along a sufficiently
long reference trajectory. Moreover, it is known that classical chaotic
systems are distinguished by their exponential sensitivity to initial
conditions and that the absence of this property in quantum systems has lead
to a number of different criteria being proposed for quantum chaos.
Exponential decay of fidelity, hypersensitivity to perturbation, and the
Zurek-Paz quantum chaos criterion of linear von Neumann's entropy growth 
\cite{zurek} are some examples \cite{caves}. These criteria accurately
predict chaos in the classical limit, but it is not clear that they behave
the same far from the classical realm.

The present work makes use of the so-called Entropic Dynamics (ED) \cite%
{caticha1}. ED is a theoretical framework that arises from the combination
of inductive inference (Maximum relative Entropy Methods, \cite{caticha2})
and Information Geometry (Riemannian geometry applied to probability theory)
(IG) \cite{amari}. As such, ED is constructed on statistical manifolds. It
is developed to investigate the possibility that laws of physics - either
classical or quantum - might reflect laws of inference rather than laws of
nature.

This article is a follow up of a series of the authors works \cite{cafaro1,
cafaro2, cafaro3}. In this paper, the ED theoretical framework is used to
explore the possibility of constructing a unified characterization of
classical and quantum chaos. We investigate a system with $3N$ degrees of
freedom (microstates), each one described by two pieces of relevant
information, its mean expected value and its variance (Gaussian statistical
macrostates). This leads to consider an ED model on a non-maximally
symmetric $6N$-dimensional statistical manifold $\mathcal{M}_{s}$. It is
shown that $\mathcal{M}_{s}$ possesses a constant negative Ricci curvature
that is proportional to the number of degrees of freedom of the system, $R_{%
\mathcal{M}_{s}}=-3N$. It is shown that the system explores statistical
volume elements on $\mathcal{M}_{s}$ at an exponential rate. We define a
dynamical information-geometric entropy $S_{\mathcal{M}_{s}}$\ of the system
and we show it increases linearly in time (statistical evolution parameter)
and is moreover, proportional to the number of degrees of freedom of the
system. The geodesics on $\mathcal{M}_{s}$ are hyperbolic trajectories.
Using the Jacobi-Levi-Civita (JLC) equation for geodesic spread, it is shown
that the Jacobi vector field intensity $J_{\mathcal{M}_{s}}$ diverges
exponentially and is proportional to the number of degrees of freedom of the
system. Thus, $R_{\mathcal{M}_{s}}$, $S_{\mathcal{M}_{s}}$ and $J_{\mathcal{M%
}_{s}}$ are proportional to the number of Gaussian-distributed microstates
of the system. This proportionality leads to conclude there is a substantial
link among these information-geometric indicators of chaoticity.

Finally, an information-geometric analog of the Zurek-Paz quantum chaos
criterion is suggested. We illustrate this point by use of an $n$-\textbf{%
set }of inverted harmonic oscillators (IHO). In the ED formalism, the IHO
system is described by a curved $n$-dimensional statistical manifold that is
conformally related to an Euclidean one.

\section{Specification of the Gaussian ED-model}

Maximum relative Entropy (ME) methods are used to construct an ED model that
follows from an assumption about what information is relevant to predict the
evolution of the system. Given a known initial macrostate (probability
distribution) and that the system evolves to a final known macrostate, the
possible trajectories of the system are examined. A notion of \textit{%
distance} between two probability distributions is provided by IG. As shown
in \cite{fisher, rao} this distance is quantified by the Fisher-Rao
information metric tensor.

We consider an ED model whose microstates span a $3N$-dimensional space
labelled by the variables $\left\{ \vec{X}\right\} =\left\{ \vec{x}^{\left(
1\right) }\text{, }\vec{x}^{\left( 2\right) }\text{,...., }\vec{x}^{\left(
N\right) }\right\} $ with $\vec{x}^{\left( \alpha \right) }\equiv \left(
x_{1}^{\left( \alpha \right) }\text{, }x_{2}^{\left( \alpha \right) }\text{, 
}x_{3}^{\left( \alpha \right) }\right) $, $\alpha =1$,...., $N$ and $%
x_{a}^{\left( \alpha \right) }\in 
%TCIMACRO{\U{211d} }%
%BeginExpansion
\mathbb{R}
%EndExpansion
$ with $a=1$, $2$, $3$. We assume the only testable information pertaining
to the quantities $x_{a}^{\left( \alpha \right) }$ consists of the
expectation values $\left\langle x_{a}^{\left( \alpha \right) }\right\rangle 
$ and variance $\Delta x_{a}^{\left( \alpha \right) }=\sqrt{\left\langle
\left( x_{a}^{\left( \alpha \right) }-\left\langle x_{a}^{\left( \alpha
\right) }\right\rangle \right) ^{2}\right\rangle }$. The set of these
expectation values define the $6N$-dimensional space of macrostates of the
system. A measure of distinguishability among the states of the ED model is
obtained by assigning a probability distribution $P\left( \vec{X}\left\vert 
\vec{\Theta}\right. \right) $ to each macrostate $\vec{\Theta}$ where $%
\left\{ \vec{\Theta}\right\} =\left\{ ^{\left( 1\right) }\theta _{a}^{\left(
\alpha \right) }\text{, }^{\left( 2\right) }\theta _{a}^{\left( \alpha
\right) }\right\} $ with $\alpha =1$, $2$,$....$, $N$ and $a=1$, $2$, $3$.
The process of assigning a probability distribution to each state endows $%
\mathcal{M}_{S}$ with a metric structure. Specifically, the Fisher-Rao
information metric defined in (\ref{fisher-rao}) is a measure of
distinguishability among macrostates. It assigns an IG to the space of
states.

\subsection{The Gaussian statistical manifold $\mathcal{M}_{S}$}

We consider an arbitrary system evolving over a $3N$-dimensional space.\ The
variables $\left\{ \vec{X}\right\} =\left\{ \vec{x}^{\left( 1\right) }\text{%
, }\vec{x}^{\left( 2\right) }\text{,...., }\vec{x}^{\left( N\right)
}\right\} $ label the $3N$-dimensional space of microstates of the system.
All information relevant to the dynamical evolution of the system is assumed
to be contained in the probability distributions. For this reason, no other
information is required. Each macrostate may be viewed as a point of a $6N$%
-dimensional statistical manifold with coordinates given by the numerical
values of the expectations $^{\left( 1\right) }\theta _{a}^{\left( \alpha
\right) }=\left\langle x_{a}^{\left( \alpha \right) }\right\rangle $ and $%
^{\left( 2\right) }\theta _{a}^{\left( \alpha \right) }=\Delta x_{a}^{\left(
\alpha \right) }\equiv \sqrt{\left\langle \left( x_{a}^{\left( \alpha
\right) }-\left\langle x_{a}^{\left( \alpha \right) }\right\rangle \right)
^{2}\right\rangle }$. The available information is contained in the
following $6N$ information constraint equations,%
\begin{equation}
\begin{array}{c}
\left\langle x_{a}^{\left( \alpha \right) }\right\rangle
=\dint\limits_{-\infty }^{+\infty }dx_{a}^{\left( \alpha \right)
}x_{a}^{\left( \alpha \right) }P_{a}^{\left( \alpha \right) }\left(
x_{a}^{\left( \alpha \right) }\left\vert ^{\left( 1\right) }\theta
_{a}^{\left( \alpha \right) }\text{,}^{\left( 2\right) }\theta _{a}^{\left(
\alpha \right) }\right. \right) \text{,} \\ 
\\ 
\Delta x_{a}^{\left( \alpha \right) }=\left[ \dint\limits_{-\infty
}^{+\infty }dx_{a}^{\left( \alpha \right) }\left( x_{a}^{\left( \alpha
\right) }-\left\langle x_{a}^{\left( \alpha \right) }\right\rangle \right)
^{2}P_{a}^{\left( \alpha \right) }\left( x_{a}^{\left( \alpha \right)
}\left\vert ^{\left( 1\right) }\theta _{a}^{\left( \alpha \right) }\text{,}%
^{\left( 2\right) }\theta _{a}^{\left( \alpha \right) }\right. \right) %
\right] ^{\frac{1}{2}}\text{,}%
\end{array}
\label{constraint1}
\end{equation}%
where $^{\left( 1\right) }\theta _{a}^{\left( \alpha \right) }=\left\langle
x_{a}^{\left( \alpha \right) }\right\rangle $ and $^{\left( 2\right) }\theta
_{a}^{\left( \alpha \right) }=\Delta x_{a}^{\left( \alpha \right) }$ with $%
\alpha =1$, $2$,$....$, $N$ and $a=1$, $2$, $3$. The probability
distributions $P_{a}^{\left( \alpha \right) }$ are constrained by the
conditions of normalization,%
\begin{equation}
\dint\limits_{-\infty }^{+\infty }dx_{a}^{\left( \alpha \right)
}P_{a}^{\left( \alpha \right) }\left( x_{a}^{\left( \alpha \right)
}\left\vert ^{\left( 1\right) }\theta _{a}^{\left( \alpha \right) }\text{,}%
^{\left( 2\right) }\theta _{a}^{\left( \alpha \right) }\right. \right) =1%
\text{.}  \label{constraint2}
\end{equation}%
The Gaussian distribution is identified by information theory as the maximum
entropy distribution if only the expectation value and the variance are
known. ME methods allows to associate a probability distribution $P\left( 
\vec{X}\left\vert \vec{\Theta}\right. \right) $ to each point in the space
of states $\vec{\Theta}$. The distribution that best reflects the
information contained in the prior distribution $m\left( \vec{X}\right) $
updated by the information $\left( \left\langle x_{a}^{\left( \alpha \right)
}\right\rangle \text{, }\Delta x_{a}^{\left( \alpha \right) }\right) $ is
obtained by maximizing the relative entropy 
\begin{equation}
S\left( \vec{\Theta}\right) =-\int\limits_{\left\{ \vec{X}\right\} }d^{3N}%
\vec{X}P\left( \vec{X}\left\vert \vec{\Theta}\right. \right) \log \left( 
\frac{P\left( \vec{X}\left\vert \vec{\Theta}\right. \right) }{m\left( \vec{X}%
\right) }\right) \text{.}  \label{entropy}
\end{equation}
As a working hypothesis, the prior $m\left( \vec{X}\right) $ is set to be
uniform since we assume the lack of prior available information about the
system (postulate of equal \textit{a priori} probabilities). Upon maximizing
(\ref{entropy}), given the constraints (\ref{constraint1}) and (\ref%
{constraint2}), we obtain%
\begin{equation}
P\left( \vec{X}\left\vert \vec{\Theta}\right. \right) =\dprod\limits_{\alpha
=1}^{N}\dprod\limits_{a=1}^{3}P_{a}^{\left( \alpha \right) }\left(
x_{a}^{\left( \alpha \right) }\left\vert \mu _{a}^{\left( \alpha \right) }%
\text{, }\sigma _{a}^{\left( \alpha \right) }\right. \right)  \label{prob}
\end{equation}%
where%
\begin{equation}
P_{a}^{\left( \alpha \right) }\left( x_{a}^{\left( \alpha \right)
}\left\vert \mu _{a}^{\left( \alpha \right) }\text{, }\sigma _{a}^{\left(
\alpha \right) }\right. \right) =\left( 2\pi \left[ \sigma _{a}^{\left(
\alpha \right) }\right] ^{2}\right) ^{-\frac{1}{2}}\exp \left[ -\frac{\left(
x_{a}^{\left( \alpha \right) }-\mu _{a}^{\left( \alpha \right) }\right) ^{2}%
}{2\left( \sigma _{a}^{\left( \alpha \right) }\right) ^{2}}\right]
\label{gaussian}
\end{equation}%
and $^{\left( 1\right) }\theta _{a}^{\left( \alpha \right) }=\mu
_{a}^{\left( \alpha \right) }$, $^{\left( 2\right) }\theta _{a}^{\left(
\alpha \right) }=\sigma _{a}^{\left( \alpha \right) }$. For the rest of the
paper, unless stated otherwise, the statistical manifold $\mathcal{M}_{S}$
will be defined by the following expression,%
\begin{equation}
\mathcal{M}_{S}=\left\{ P\left( \vec{X}\left\vert \vec{\Theta}\right.
\right) \text{ in (\ref{prob})}:\vec{X}\in 
%TCIMACRO{\U{211d} }%
%BeginExpansion
\mathbb{R}
%EndExpansion
^{3N}\text{, }\vec{\Theta}\in \mathcal{D}_{\Theta }=\left[ \left( -\infty 
\text{, }+\infty \right) _{\mu }\times \left( 0\text{, }+\infty \right)
_{\sigma }\right] ^{3N}\right\} \text{.}  \label{manifold}
\end{equation}%
The probability distribution (\ref{prob}) encodes the available information
concerning the system. Note we assumed uncoupled constraints among
microvariables $x_{a}^{\left( \alpha \right) }$. In other words, we assumed
that information about correlations between the microvariables need not to
be tracked. This assumption leads to the simplified product rule (\ref{prob}%
). However, coupled constraints would lead to a generalized product rule in (%
\ref{prob}) and to a metric tensor (\ref{fisher-rao}) with non-trivial
off-diagonal elements (covariance terms). For instance, the total
probability distribution $P\left( x\text{, }y|\mu _{x}\text{, }\sigma _{x}%
\text{, }\mu _{y}\text{, }\sigma _{y}\right) $ of two dependent Gaussian
distributed microvariables $x$ and $y$ reads%
\begin{eqnarray}
&&P\left( x\text{, }y|\mu _{x}\text{, }\sigma _{x}\text{, }\mu _{y}\text{, }%
\sigma _{y}\right) =\frac{1}{2\pi \sigma _{x}\sigma _{y}\sqrt{1-r^{2}}}\times
\label{corr-prob} \\
&&\times \exp \left\{ -\frac{1}{2\left( 1-r^{2}\right) }\left[ \frac{\left(
x-\mu _{x}\right) ^{2}}{\sigma _{x}^{2}}-2r\frac{\left( x-\mu _{x}\right)
\left( y-\mu _{y}\right) }{\sigma _{x}\sigma _{y}}+\frac{\left( y-\mu
_{y}\right) ^{2}}{\sigma _{y}^{2}}\right] \right\} \text{,}  \notag
\end{eqnarray}%
where $r\in \left( -1\text{, }+1\right) $ is the correlation coefficient
given by%
\begin{equation}
r=\frac{\left\langle \left( x-\left\langle x\right\rangle \right) \left(
y-\left\langle y\right\rangle \right) \right\rangle }{\sqrt{\left\langle
x-\left\langle x\right\rangle \right\rangle }\sqrt{\left\langle
y-\left\langle y\right\rangle \right\rangle }}=\frac{\left\langle
xy\right\rangle -\left\langle x\right\rangle \left\langle y\right\rangle }{%
\sigma _{x}\sigma _{y}}\text{.}
\end{equation}%
The metric induced by (\ref{corr-prob}) is obtained by use of (\ref%
{fisher-rao}), the result being%
\begin{equation}
g_{ij}=\left[ 
\begin{array}{cccc}
-\frac{1}{\sigma _{x}^{2}\left( r^{2}-1\right) } & 0 & \frac{r}{\sigma
_{x}\sigma _{y}\left( r^{2}-1\right) } & 0 \\ 
0 & -\frac{2-r^{2}}{\sigma _{x}^{2}\left( r^{2}-1\right) } & 0 & \frac{r^{2}%
}{\sigma _{x}\sigma _{y}\left( r^{2}-1\right) } \\ 
\frac{r}{\sigma _{x}\sigma _{y}\left( r^{2}-1\right) } & 0 & -\frac{1}{%
\sigma _{y}^{2}\left( r^{2}-1\right) } & 0 \\ 
0 & \frac{r^{2}}{\sigma _{x}\sigma _{y}\left( r^{2}-1\right) } & 0 & -\frac{%
2-r^{2}}{\sigma _{y}^{2}\left( r^{2}-1\right) }%
\end{array}%
\right] \text{,}  \label{corr-metric}
\end{equation}%
where $i$, $j=1$, $2$, $3$, $4$. The Ricci curvature scalar associated with
manifold characterized by (\ref{corr-metric}) is given by%
\begin{equation}
R=g^{ij}R_{ij}=-\frac{8\left( r^{2}-2\right) +2r^{2}\left( 3r^{2}-2\right) }{%
8\left( r^{2}-1\right) }\text{.}
\end{equation}%
It is clear that in the limit $r\rightarrow 0$, the off-diagonal elements of 
$g_{ij}$ vanish and the Scalar $R$ reduces to the result obtained in \cite%
{cafaro2}, namely $R=-2<0$. Correlation terms may be fictitious. They may
arise for instance from coordinate transformations. On the other hand,
correlations may arise from external fields in which the system is immersed.
In such situations, correlations among $x_{a}^{\left( \alpha \right) }$
effectively describe interaction between the microvariables and the external
fields. Such generalizations would require more delicate analysis. Before
proceeding, a comment is in order. Most probability distributions arise from
the maximum entropy formalism as a result of simple statements concerning
averages (Gaussians, exponential, binomial, etc.). Not all distribution are
generated in this manner however. Some distributions are generated by
combining the results of simple cases (multinomial from a binomial) while
others are found as a result of a change of variables (Cauchy distribution).
For instance, the Weibull and Wigner-Dyson distributions can be obtained
from an exponential distribution as a result of a power law transformation 
\cite{wigner-dyson}.

\subsubsection{Metric structure of $\mathcal{M}_{S}$}

We cannot determine the evolution of microstates of the system since the
available information is insufficient. Not only is the information available
insufficient but we also do not know the equation of motion. In fact there
is no standard "equation of motion".\ Instead we can ask: how close are the
two total distributions with parameters $(\mu _{a}^{\left( \alpha \right) }$%
, $\sigma _{a}^{\left( \alpha \right) })$ and $(\mu _{a}^{\left( \alpha
\right) }+d\mu _{a}^{\left( \alpha \right) }$, $\sigma _{a}^{\left( \alpha
\right) }+d\sigma _{a}^{\left( \alpha \right) })$? Once the states of the
system have been defined, the next step concerns the problem of quantifying
the notion of change from the state $\vec{\Theta}$ to the state $\vec{\Theta}%
+d\vec{\Theta}$. A convenient measure of change is distance. The measure we
seek is given by the dimensionless \textit{distance} $ds$ between $P\left( 
\vec{X}\left\vert \vec{\Theta}\right. \right) $ and $P\left( \vec{X}%
\left\vert \vec{\Theta}+d\vec{\Theta}\right. \right) $,%
\begin{equation}
ds^{2}=g_{\mu \nu }d\Theta ^{\mu }d\Theta ^{\nu }\text{ with }\mu \text{, }%
\nu =1\text{, }2\text{,.., }6N\text{,}  \label{line-element}
\end{equation}%
where%
\begin{equation}
g_{\mu \nu }=\int d\vec{X}P\left( \vec{X}\left\vert \vec{\Theta}\right.
\right) \frac{\partial \log P\left( \vec{X}\left\vert \vec{\Theta}\right.
\right) }{\partial \Theta ^{\mu }}\frac{\partial \log P\left( \vec{X}%
\left\vert \vec{\Theta}\right. \right) }{\partial \Theta ^{\nu }}
\label{fisher-rao}
\end{equation}%
is the Fisher-Rao information metric. Substituting (\ref{prob}) into (\ref%
{fisher-rao}), the metric $g_{\mu \nu }$ on $\mathcal{M}_{s}$ becomes a $%
6N\times 6N$ matrix $M$ made up of $3N$ blocks $M_{2\times 2}$ with
dimension $2\times 2$ given by,%
\begin{equation}
M_{2\times 2}=\left( 
\begin{array}{cc}
\left( \sigma _{a}^{\left( \alpha \right) }\right) ^{-2} & 0 \\ 
0 & 2\times \left( \sigma _{a}^{\left( \alpha \right) }\right) ^{-2}%
\end{array}%
\right)  \label{m-matrix}
\end{equation}%
with $\alpha =1$, $2$,$....$, $N$ and $a=1,2,3$. From (\ref{fisher-rao}),
the "length" element (\ref{line-element}) reads,%
\begin{equation}
ds^{2}=\dsum\limits_{\alpha =1}^{N}\dsum\limits_{a=1}^{3}\left[ \frac{1}{%
\left( \sigma _{a}^{\left( \alpha \right) }\right) ^{2}}d\mu _{a}^{\left(
\alpha \right) 2}+\frac{2}{\left( \sigma _{a}^{\left( \alpha \right)
}\right) ^{2}}d\sigma _{a}^{\left( \alpha \right) 2}\right] \text{.}
\end{equation}%
We bring attention to the fact that the metric structure of $\mathcal{M}_{s}$
is an emergent (not fundamental) structure. It arises only after assigning a
probability distribution $P\left( \vec{X}\left\vert \vec{\Theta}\right.
\right) $ to each state $\vec{\Theta}$.

\subsubsection{Curvature of\ $\mathcal{M}_{s}$}

Given the Fisher-Rao information metric, we use standard differential
geometry methods applied to the space of probability distributions to
characterize the geometric properties of $\mathcal{M}_{s}$. Recall that the
Ricci scalar curvature $R$ is given by,%
\begin{equation}
R=g^{\mu \nu }R_{\mu \nu }\text{,}  \label{ricci-scalar}
\end{equation}%
where $g^{\mu \nu }g_{\nu \rho }=\delta _{\rho }^{\mu }$ so that $g^{\mu \nu
}=\left( g_{\mu \nu }\right) ^{-1}$. The Ricci tensor $R_{\mu \nu }$ is
given by,%
\begin{equation}
R_{\mu \nu }=\partial _{\gamma }\Gamma _{\mu \nu }^{\gamma }-\partial _{\nu
}\Gamma _{\mu \lambda }^{\lambda }+\Gamma _{\mu \nu }^{\gamma }\Gamma
_{\gamma \eta }^{\eta }-\Gamma _{\mu \gamma }^{\eta }\Gamma _{\nu \eta
}^{\gamma }\text{.}
\end{equation}%
The Christoffel symbols $\Gamma _{\mu \nu }^{\rho }$ appearing in the Ricci
tensor are defined in the standard manner as, 
\begin{equation}
\Gamma _{\mu \nu }^{\rho }=\frac{1}{2}g^{\rho \sigma }\left( \partial _{\mu
}g_{\sigma \nu }+\partial _{\nu }g_{\mu \sigma }-\partial _{\sigma }g_{\mu
\nu }\right) .  \label{christoffel}
\end{equation}%
Using (\ref{m-matrix}) and the definitions given above, we can show that the
Ricci scalar curvature becomes%
\begin{equation}
R_{\mathcal{M}_{s}}=R_{\text{ }\alpha }^{\alpha }=\sum_{\rho \neq \sigma
}K\left( e_{\rho }\text{, }e_{\sigma }\right) =-3N<0\text{.}  \label{Ricci}
\end{equation}%
The scalar curvature is the sum of all sectional curvatures of planes
spanned by pairs of orthonormal basis elements $\left\{ e_{\rho }=\partial
_{\Theta _{\rho }(p)}\right\} $ of the tangent space $T_{p}\mathcal{M}_{s}$
with $p\in \mathcal{M}_{s}$, 
\begin{equation}
K\left( a\text{, }b\right) =\frac{R_{\mu \nu \rho \sigma }a^{\mu }b^{\nu
}a^{\rho }b^{\sigma }}{\left( g_{\mu \sigma }g_{\nu \rho }-g_{\mu \rho
}g_{\nu \sigma }\right) a^{\mu }b^{\nu }a^{\rho }b^{\sigma }}\text{, }%
a=\sum_{\rho }\left\langle a\text{, }h^{\rho }\right\rangle e_{\rho }\text{,}
\label{sectionK}
\end{equation}%
where $\left\langle e_{\rho }\text{, }h^{\sigma }\right\rangle =\delta
_{\rho }^{\sigma }$. Notice that the sectional curvatures completely
determine the curvature tensor. From (\ref{Ricci}) we conclude that $%
\mathcal{M}_{s}$ is a $6N$-dimensional statistical manifold of constant
negative Ricci scalar curvature. A detailed analysis on the calculation of
Christoffel connection coefficients using the ED formalism for a
four-dimensional manifold of Gaussians can be found in \cite{cafaro2}.

\subsubsection{Anisotropy and Compactness}

It can be shown that $\mathcal{M}_{s}$ is not a pseudosphere (maximally
symmetric manifold). The first way this can be understood is from the fact
that the Weyl Projective curvature tensor \cite{goldberg} (or the anisotropy
tensor) $W_{\mu \nu \rho \sigma }$ defined by%
\begin{equation}
W_{\mu \nu \rho \sigma }=R_{\mu \nu \rho \sigma }-\frac{R_{\mathcal{M}_{s}}}{%
n\left( n-1\right) }\left( g_{\nu \sigma }g_{\mu \rho }-g_{\nu \rho }g_{\mu
\sigma }\right) \text{,}  \label{Weyl}
\end{equation}%
with $n=6N$ in the present case, is non-vanishing. In (\ref{Weyl}), the
quantity $R_{\mu \nu \rho \sigma }$ is the Riemann curvature tensor defined
in the usual manner by%
\begin{equation}
R^{\alpha }\,_{\beta \rho \sigma }=\partial _{\sigma }\Gamma _{\text{ \ }%
\beta \rho }^{\alpha }-\partial _{\rho }\Gamma _{\text{ \ }\beta \sigma
}^{\alpha }+\Gamma ^{\alpha }\,_{\lambda \sigma }\Gamma ^{\lambda }\,_{\beta
\rho }-\Gamma ^{\alpha }\,_{\lambda \rho }\Gamma ^{\lambda }\,_{\beta \sigma
}\text{.}
\end{equation}%
Considerations regarding the negativity of the Ricci curvature as a \textit{%
strong criterion} of dynamical instability and the necessity of \textit{%
compactness} of $\mathcal{M}_{s}$\textit{\ }in "true" chaotic dynamical
systems is under investigation \cite{cafaro4}.

The issue of symmetry of $\mathcal{M}_{s}$ can alternatively be understood
from consideration of the sectional curvature. In view of (\ref{sectionK}),
the negativity of the Ricci scalar implies the existence of expanding
directions in the configuration space manifold $\mathcal{M}_{s}$. Indeed,
from (\ref{Ricci}) one may conclude that negative principal curvatures
(extrema of sectional curvatures) dominate over positive ones. Thus, the
negativity of the Ricci scalar is only a \textit{sufficient} (not necessary)
condition for local instability of geodesic flow. For this reason, the
negativity of the scalar provides a \textit{strong }criterion of local
instability. Scenarios may arise where negative sectional curvatures are
present, but the positive ones could prevail in the sum so that the Ricci
scalar is non-negative despite the instability in the flow in those
directions. Consequently, the signs of the sectional curvatures are of
primary significance for the proper characterization of chaos.

Yet another useful way to understand the anisotropy of the $\mathcal{M}_{s}$
is the following. It is known that in $n$ dimensions, there are at most $%
\frac{n\left( n+1\right) }{2}$ independent Killing vectors (directions of
symmetry of the manifold). Since $\mathcal{M}_{s}$ is not a pseudosphere,
the information metric tensor does not admit the maximum number of Killing
vectors $K_{\nu }$ defined as%
\begin{equation}
\mathcal{L}_{K}g_{\mu \nu }=D_{\mu }K_{\nu }+D_{\nu }K_{\mu }=0\text{,}
\end{equation}%
where $D_{\mu }$, given by%
\begin{equation}
D_{\mu }K_{\nu }=\partial _{\mu }K_{\nu }-\Gamma _{\nu \mu }^{\rho }K_{\rho }%
\text{,}
\end{equation}%
is the covariant derivative operator with respect to the connection $\Gamma $
defined in (\ref{christoffel}). The Lie derivative $\mathcal{L}_{K}g_{\mu
\nu }$ of the tensor field $g_{\mu \nu }$ along a given direction $K$
measures the intrinsic variation of the field along that direction (that is,
the metric tensor is Lie transported along the Killing vector) \cite{clarke}%
. Locally, a maximally symmetric space of Euclidean signature is either a
plane, a sphere, or a hyperboloid, depending on the sign of $R$. In our
case, none of these scenarios occur. As will be seen in what follows, this
fact has a significant impact on the integration of the geodesic deviation
equation on $\mathcal{M}_{s}$. At this juncture, we emphasize it is known
that the anisotropy of the manifold underlying system dynamics plays a
crucial role in the mechanism of instability. In particular, fluctuating
sectional curvatures require also that the manifold be anisotropic. However,
the connection between curvature variations along geodesics and anisotropy
is far from clear and is currently under investigation.

Krylov was the first to emphasize \cite{krylov} the use of $R<0$ as an
instability criterion in the context of an $N$-body system (a gas)
interacting via Van der Waals forces, with the ultimate hope to understand
the relaxation process in a gas. However, Krylov neglected the problem of
compactness of the configuration space manifold which is important for
making inferences about exponential mixing of geodesic flows \cite{pellicott}%
. Why is compactness so significant in the characterization of chaos? 
\textit{True} chaos should be identified by the occurrence of two crucial
features: 1) strong dependence on initial conditions and exponential
divergence of the Jacobi vector field intensity, i.e., \textit{stretching}
of dynamical trajectories; 2) compactness of the configuration space
manifold, i.e., \textit{folding} of dynamical trajectories. Compactness \cite%
{cipriani, jost} is required in order to discard trivial exponential growths
due to the unboundedness of the "volume" available to the dynamical system.
In other words, the folding is necessary to have a dynamics actually able to
mix the trajectories, making practically impossible, after a finite interval
of time, to discriminate between trajectories which were very nearby each
other at the initial time. When the space is not compact, even in presence
of strong dependence on initial conditions, it could be possible in some
instances (though not always), to distinguish among different trajectories
originating within a small distance and then evolved subject to exponential
instability.

The statistical manifold defined in (\ref{manifold}) is compact. This can be
seen as follows. It is known form IG that there is a one-to-one relation
between elements of the statistical manifold and the parameter space. More
precisely, the statistical manifold $\mathcal{M}_{s}$ is \textit{homeomorphic%
} to the parameter space $\mathcal{D}_{\Theta }$. This implies the existence
of a continuous, bijective map $h_{\mathcal{M}_{s}\text{, }\mathcal{D}%
_{\Theta }}$,%
\begin{equation}
h_{\mathcal{M}_{s}\text{, }\mathcal{D}_{\Theta }\text{ }}:\mathcal{M}_{S}\ni
P\left( \vec{X}\left\vert \vec{\Theta}\right. \right) \rightarrow \vec{\Theta%
}\in \mathcal{D}_{\Theta }
\end{equation}%
where $h_{\mathcal{M}_{s}\text{, }\mathcal{D}_{\Theta }\text{ }}^{-1}\left( 
\vec{\Theta}\right) =P\left( \vec{X}\left\vert \vec{\Theta}\right. \right) $%
. The inverse image $h_{\mathcal{M}_{s}\text{, }\mathcal{D}_{\Theta }\text{ }%
}^{-1}$ is the so-called homeomorphism map. In addition, since
homeomorphisms preserve compactness, it is sufficient to restrict ourselves
to a compact subspace of the parameter space $\mathcal{D}_{\Theta }$ in
order to ensure that $\mathcal{M}_{S}$ is itself compact.

\section{Canonical formalism for the Gaussian ED-model}

The geometrization of a Hamiltonian system by transforming it to a geodesic
flow is a well-known technique of classical mechanics associated with the
name of Jacobi \cite{jacobi}. Transformation to geodesic motion is obtained
in two steps: 1) conformal transformation of the metric; 2) rescaling of the
time parameter \cite{biesiada1}. The reformulation of dynamics in terms of a
geodesic problem allows the application of a wide range of well-known
geometrical techniques in the investigation of the solution space and
properties of equations of motions. The power of the Jacobi reformulation is
that all of the dynamical information is collected into a single geometric
object - the manifold on which geodesic flow is induced - in which all the
available manifest symmetries are retained. For instance, integrability of
the system is connected with the existence of Killing vectors and tensors on
this manifold \cite{biesiada2, uggla}.

In this section we study the trajectories of the system on $\mathcal{M}_{s}$%
. We emphasize ED can be derived from a standard principle of least action
(of Maupertuis-Euler-Lagrange-Jacobi type) \cite{caticha1, arnold}. The main
differences are that the dynamics being considered here, namely ED, is
defined on a space of probability distributions $\mathcal{M}_{s}$, not on an
ordinary linear space $V$ and the standard coordinates $q_{\mu }$ of the
system are replaced by statistical macrovariables $\Theta ^{\mu }$. The
geodesic equations for the macrovariables of the Gaussian ED model are given
by,%
\begin{equation}
\frac{d^{2}\Theta ^{\mu }}{d\tau ^{2}}+\Gamma _{\nu \rho }^{\mu }\frac{%
d\Theta ^{\nu }}{d\tau }\frac{d\Theta ^{\rho }}{d\tau }=0  \label{geodesic}
\end{equation}%
with $\mu =1$, $2$,..., $6N$. Observe the geodesic equations are\textit{\
nonlinear} second order coupled ordinary differential equations. They
describe a \textit{reversible} dynamics whose solution is the trajectory
between an initial and a final macrostate. The trajectory can be equally
well traversed in both directions.

\subsection{Geodesics on $\mathcal{M}_{s}$}

We determine the explicit form of (\ref{geodesic}) for the pairs of
statistical coordinates $(\mu _{a}^{\left( \alpha \right) }$, $\sigma
_{a}^{\left( \alpha \right) })$. Substituting the expression of the
Christoffel connection coefficients into (\ref{geodesic}), the geodesic
equations for the macrovariables $\mu _{a}^{\left( \alpha \right) }$ and $%
\sigma _{a}^{\left( \alpha \right) }$ associated to the microstate $%
x_{a}^{\left( \alpha \right) }$ become,%
\begin{equation}
\frac{d^{2}\mu _{a}^{\left( \alpha \right) }}{d\tau ^{2}}-\frac{2}{\sigma
_{a}^{\left( \alpha \right) }}\frac{d\mu _{a}^{\left( \alpha \right) }}{%
d\tau }\frac{d\sigma _{a}^{\left( \alpha \right) }}{d\tau }=0\text{, }\frac{%
d^{2}\sigma _{a}^{\left( \alpha \right) }}{d\tau ^{2}}-\frac{1}{\sigma
_{a}^{\left( \alpha \right) }}\left( \frac{d\sigma _{a}^{\left( \alpha
\right) }}{d\tau }\right) ^{2}+\frac{1}{2\sigma _{a}^{\left( \alpha \right) }%
}\left( \frac{d\mu _{a}^{\left( \alpha \right) }}{d\tau }\right) ^{2}=0\text{%
,}
\end{equation}%
with $\alpha =1$, $2$,$....$, $N$ and $a=1,2,3$. This is a set of coupled
ordinary differential equations, whose solutions are%
\begin{equation}
\begin{array}{c}
\mu _{a}^{\left( \alpha \right) }\left( \tau \right) =\frac{\left(
B_{a}^{\left( \alpha \right) }\right) ^{2}}{2\beta _{a}^{\left( \alpha
\right) }}\frac{1}{\cosh \left( 2\beta _{a}^{\left( \alpha \right) }\tau
\right) -\sinh \left( 2\beta _{a}^{\left( \alpha \right) }\tau \right) +%
\frac{\left( B_{a}^{\left( \alpha \right) }\right) ^{2}}{8\left( \beta
_{a}^{\left( \alpha \right) }\right) ^{2}}}+C_{a}^{\left( \alpha \right) }%
\text{,} \\ 
\\ 
\sigma _{a}^{\left( \alpha \right) }\left( \tau \right) =B_{a}^{\left(
\alpha \right) }\frac{\cosh \left( \beta _{a}^{\left( \alpha \right) }\tau
\right) -\sinh \left( \beta _{a}^{\left( \alpha \right) }\tau \right) }{%
\cosh \left( 2\beta _{a}^{\left( \alpha \right) }\tau \right) -\sinh \left(
2\beta _{a}^{\left( \alpha \right) }\tau \right) +\frac{\left( B_{a}^{\left(
\alpha \right) }\right) ^{2}}{8\left( \beta _{a}^{\left( \alpha \right)
}\right) ^{2}}}\text{.}%
\end{array}%
\end{equation}%
The quantities $B_{a}^{\left( \alpha \right) }$, $C_{a}^{\left( \alpha
\right) }$, $\beta _{a}^{\left( \alpha \right) }$ are \textit{real}
integration constants that can be evaluated upon specification of boundary
conditions. We are interested in the stability of the trajectories on $%
\mathcal{M}_{s}$. It is known \cite{arnold} that the Riemannian curvature of
a manifold is intimately related to the behavior of geodesics on it. If the
Riemannian curvature of a manifold is negative, geodesics (initially
parallel) rapidly diverge from one another. For the sake of simplicity, we
assume very special initial conditions: $B_{a}^{\left( \alpha \right)
}\equiv \Lambda $, $\beta _{a}^{\left( \alpha \right) }\equiv \lambda \in 
%TCIMACRO{\U{211d} }%
%BeginExpansion
\mathbb{R}
%EndExpansion
^{+}$, $C_{a}^{\left( \alpha \right) }=0$, $\forall \alpha =1$, $2$,$....$, $%
N$ and $a=1$, $2$, $3$. However, the conclusions drawn can be generalized to
more arbitrary initial conditions. We observe that since every maximal
geodesic is well-defined for all temporal parameters $\tau $, $\mathcal{M}%
_{s}$ constitute a geodesically complete manifold \cite{lee}. It is
therefore a natural setting within which one may consider \textit{global}
questions and search for a \textit{weak criterion} of chaos \cite{cipriani}.

\section{Exponential divergence of the Jacobi vector field intensity}

The actual interest of the Riemannian formulation of the dynamics stems form
the possibility of studying the instability of natural motions through the
instability of geodesics of a suitable manifold, a circumstance that has
several advantages. First of all a powerful mathematical tool exists to
investigate the stability or instability of a geodesic flow: the
Jacobi-Levi-Civita equation for geodesic spread \cite{carmo}. The
JLC-equation describes covariantly how nearby geodesics locally scatter. It
is a familiar object both in Riemannian geometry and theoretical physics (it
is of fundamental interest in experimental General Relativity). Moreover the
JLC-equation relates the stability or instability of a geodesic flow with
curvature properties of the ambient manifold, thus opening a wide and
largely unexplored field of investigation of the connections among geometry,
topology and geodesic instability, hence chaos.

Consider the behavior of the one-parameter family of neighboring geodesics $%
\mathcal{F}_{G_{\mathcal{M}_{s}}}\left( \lambda \right) \equiv \left\{
\Theta _{\mathcal{M}_{s}}^{\mu }\left( \tau \text{; }\lambda \right)
\right\} _{\lambda \in 
%TCIMACRO{\U{211d} }%
%BeginExpansion
\mathbb{R}
%EndExpansion
^{+}}^{\mu =1\text{,.., }6N}$ where%
\begin{eqnarray}
\mu _{a}^{\left( \alpha \right) }\left( \tau \text{; }\lambda \right) &=&%
\frac{\Lambda ^{2}}{2\lambda }\frac{1}{\cosh \left( 2\lambda \tau \right)
-\sinh \left( 2\lambda \tau \right) +\frac{\Lambda ^{2}}{8\lambda ^{2}}}%
\text{,}  \notag \\
&&  \label{solns} \\
\sigma _{a}^{\left( \alpha \right) }\left( \tau \text{; }\lambda \right)
&=&\Lambda \frac{\cosh \left( \lambda \tau \right) -\sinh \left( \lambda
\tau \right) }{\cosh \left( 2\lambda \tau \right) -\sinh \left( 2\lambda
\tau \right) +\frac{\Lambda ^{2}}{8\lambda ^{2}}}\text{.}  \notag
\end{eqnarray}%
with $\alpha =1$, $2$,$....$, $N$ and $a=1$, $2$, $3$. The relative geodesic
spread on a (non-maximally symmetric) curved manifold as $\mathcal{M}_{s}$
is characterized by the Jacobi-Levi-Civita equation, the natural tool to
tackle dynamical chaos \cite{clarke, carmo},%
\begin{equation}
\frac{D^{2}\delta \Theta ^{\mu }}{D\tau ^{2}}+R_{\nu \rho \sigma }^{\mu }%
\frac{\partial \Theta ^{\nu }}{\partial \tau }\delta \Theta ^{\rho }\frac{%
\partial \Theta ^{\sigma }}{\partial \tau }=0  \label{gen-geoDev}
\end{equation}%
where the Jacobi vector field $J^{\mu }$ is defined as,%
\begin{equation}
J^{\mu }\equiv \delta \Theta ^{\mu }\overset{\text{def}}{=}\delta _{\lambda
}\Theta ^{\mu }=\left. \left( \frac{\partial \Theta ^{\mu }\left( \tau \text{%
; }\lambda \right) }{\partial \lambda }\right) \right\vert _{\tau =\text{%
const}}\delta \lambda \text{.}  \label{jacobi}
\end{equation}%
Notice that the JLC-equation appears intractable already at rather small $N$%
. For isotropic manifolds, the JLC-equation can be reduced to the simple
form,%
\begin{equation}
\frac{D^{2}J^{\mu }}{D\tau ^{2}}+KJ^{\mu }=0\text{, }\mu =1\text{,...., }6N
\label{geo-deviation}
\end{equation}%
where $K$ is the constant value assumed throughout the manifold by the
sectional curvature. The sectional curvature of manifold $\mathcal{M}_{s}$
is the $6N$-dimensional generalization of the Gaussian curvature of
two-dimensional surfaces of $%
%TCIMACRO{\U{211d} }%
%BeginExpansion
\mathbb{R}
%EndExpansion
^{3}$. If $K<0$, unstable solutions of equation (\ref{geo-deviation})
assumes the form%
\begin{equation}
J\left( \tau \right) =\frac{1}{\sqrt{-K}}\omega \left( 0\right) \sinh \left( 
\sqrt{-K}\tau \right)
\end{equation}%
once the initial conditions are assigned as $J\left( 0\right) =0$, $\frac{%
dJ\left( 0\right) }{d\tau }=\omega \left( 0\right) $ and $K<0$. Equation (%
\ref{gen-geoDev}) forms a system of $6N$ coupled ordinary differential
equations \textit{linear} in the components of the deviation vector field (%
\ref{jacobi}) but\textit{\ nonlinear} in derivatives of the metric (\ref%
{fisher-rao}). It describes the linearized geodesic flow: the linearization
ignores the relative velocity of the geodesics. When the geodesics are
neighboring but their relative velocity is arbitrary, the corresponding
geodesic deviation equation is the so-called generalized Jacobi equation 
\cite{chicone, hodgkinson}. The nonlinearity is due to the existence of
velocity-dependent terms in the system. Neighboring geodesics accelerate
relative to each other with a rate directly measured by the curvature tensor 
$R_{\alpha \beta \gamma \delta }$. Substituting (\ref{solns}) in (\ref%
{gen-geoDev}) and neglecting the exponentially decaying terms in $\delta
\Theta ^{\mu }$ and its derivatives, integration of (\ref{gen-geoDev}) leads
to the following asymptotic expression of the Jacobi vector field intensity,%
\begin{equation}
J_{\mathcal{M}_{S}}=\left\Vert J\right\Vert =\left( g_{\mu \nu }J^{\mu
}J^{\nu }\right) ^{\frac{1}{2}}\overset{\tau \rightarrow \infty }{\approx }%
3Ne^{\lambda \tau }\text{.}
\end{equation}%
We conclude that the geodesic spread on $\mathcal{M}_{s}$ is described by
means of an \textit{exponentially} \textit{divergent} Jacobi vector field
intensity $J_{\mathcal{M}_{s}}$, a \textit{classical} feature of chaos. In
our approach the quantity $\lambda _{J}$,%
\begin{equation}
\lambda _{J}\overset{\text{def}}{=}\underset{\tau \rightarrow \infty }{\lim }%
\frac{1}{\tau }\ln \left[ \frac{\left\Vert J_{_{\mathcal{M}_{S}}}\left( \tau
\right) \right\Vert }{\left\Vert J_{_{\mathcal{M}_{S}}}\left( 0\right)
\right\Vert }\right]
\end{equation}%
would play the role of the conventional Lyapunov exponents.

\section{Linearity of the information geometrodynamical entropy}

We investigate the stability of the trajectories of the ED model considered
on $\mathcal{M}_{s}$. It is known \cite{arnold} that the Riemannian
curvature of a manifold is closely connected with the behavior of the
geodesics on it. If the Riemannian curvature of a manifold is negative,
geodesics (initially parallel) rapidly diverge from one another. For the
sake of simplicity, we assume very special initial conditions: $%
B_{a}^{\left( \alpha \right) }\equiv \Lambda $, $\beta _{a}^{\left( \alpha
\right) }\equiv \lambda \in 
%TCIMACRO{\U{211d} }%
%BeginExpansion
\mathbb{R}
%EndExpansion
^{+}$, $C_{a}^{\left( \alpha \right) }=0$, $\forall $ $\alpha =1$, $2$,$....$%
, $N$ and $a=1$, $2$, $3$ . However, the conclusion we reach can be
generalized to more arbitrary initial conditions. Recall $\mathcal{M}_{s}$
is the space of probability distributions $\left\{ P\left( \vec{X}\left\vert 
\vec{\Theta}\right. \right) \right\} $ labeled by $6N$ statistical
parameters $\vec{\Theta}$. These parameters are the coordinates for the
point $P$, and in these coordinates a volume element $dV_{\mathcal{M}_{s}}$
reads, 
\begin{equation}
dV_{\mathcal{M}_{S}}=\sqrt{g}d^{6N}\vec{\Theta}=\dprod\limits_{\alpha
=1}^{N}\dprod\limits_{a=1}^{3}\frac{\sqrt{2}}{\left( \sigma _{a}^{\left(
\alpha \right) }\right) ^{2}}d\mu _{a}^{\left( \alpha \right) }d\sigma
_{a}^{\left( \alpha \right) }\text{.}
\end{equation}%
The volume of an extended region $\Delta V_{\mathcal{M}_{s}}\left( \tau 
\text{; }\lambda \right) $ of $\mathcal{M}_{s}$ is defined by,%
\begin{equation}
\Delta V_{\mathcal{M}_{s}}\left( \tau \text{; }\lambda \right) \overset{%
\text{def}}{=}\dprod\limits_{\alpha
=1}^{N}\dprod\limits_{a=1}^{3}\int\nolimits_{\mu _{a}^{\left( \alpha \right)
}\left( 0\right) }^{\mu _{a}^{\left( \alpha \right) }\left( \tau \right)
}\int\nolimits_{\sigma _{a}^{\left( \alpha \right) }\left( 0\right)
}^{\sigma _{a}^{\left( \alpha \right) }\left( \tau \right) }\frac{\sqrt{2}}{%
\left( \sigma _{a}^{\left( \alpha \right) }\right) ^{2}}d\mu _{a}^{\left(
\alpha \right) }d\sigma _{a}^{\left( \alpha \right) }
\end{equation}%
where $\mu _{a}^{\left( \alpha \right) }\left( \tau \right) $ and $\sigma
_{a}^{\left( \alpha \right) }\left( \tau \right) $ are given in (\ref{solns}%
). The quantity that encodes relevant information about the stability of
neighboring volume elements is the the average volume $\bar{V}_{\mathcal{M}%
_{s}}\left( \tau \text{; }\lambda \right) $, 
\begin{equation}
\bar{V}_{\mathcal{M}_{s}}\left( \tau \text{; }\lambda \right) \equiv \left.
\left\langle \Delta V_{\mathcal{M}_{s}}\left( \tau \text{; }\lambda \right)
\right\rangle \right\vert _{\tau }\overset{\text{def}}{=}\frac{1}{\tau }%
\dint\limits_{0}^{\tau }\Delta V_{\mathcal{M}_{s}}\left( \tau ^{\prime }%
\text{; }\lambda \right) d\tau ^{\prime }\overset{\tau \rightarrow \infty }{%
\approx }e^{3N\lambda \tau }\text{.}  \label{avg-vol}
\end{equation}%
This asymptotic regime of diffusive evolution in (\ref{avg-vol}) describes
the exponential increase of average volume elements on $\mathcal{M}_{s}$.
The exponential instability characteristic of chaos forces the system to
rapidly explore large areas (volumes) of the statistical manifold. It is
interesting to note that this asymptotic behavior appears also in the
conventional description of quantum chaos where the entropy increases
linearly at a rate determined by the Lyapunov exponents \cite{ruelle}. The
linear increase of entropy as a quantum chaos criterion was introduced by
Zurek and Paz \cite{zurek}. In our information-geometric approach a relevant
quantity that can be useful to study the degree of instability
characterizing the ED model is the information-geometric entropy defined as,%
\begin{equation}
S_{\mathcal{M}_{s}}\overset{\text{def}}{=}\underset{\tau \rightarrow \infty }%
{\lim }\log \bar{V}_{\mathcal{M}_{s}}\left( \tau \text{; }\lambda \right) 
\text{.}  \label{asym-ent}
\end{equation}%
Substituting (\ref{avg-vol}) in (\ref{asym-ent}), we obtain%
\begin{equation}
S_{\mathcal{M}_{s}}=\underset{\tau \rightarrow \infty }{\lim }\log \left\{ 
\frac{1}{\tau }\dint\limits_{0}^{\tau }\left[ \dprod\limits_{\alpha
=1}^{N}\dprod\limits_{a=1}^{3}\int\nolimits_{\mu _{a}^{\left( \alpha \right)
}\left( 0\right) }^{\mu _{a}^{\left( \alpha \right) }\left( \tau ^{\prime
}\right) }\int\nolimits_{\sigma _{a}^{\left( \alpha \right) }\left( 0\right)
}^{\sigma _{a}^{\left( \alpha \right) }\left( \tau ^{\prime }\right) }\frac{%
\sqrt{2}}{\left( \sigma _{a}^{\left( \alpha \right) }\right) ^{2}}d\mu
_{a}^{\left( \alpha \right) }d\sigma _{a}^{\left( \alpha \right) }\right]
d\tau ^{\prime }\right\} \overset{\tau \rightarrow \infty }{\approx }%
3N\lambda \tau \text{.}  \label{ent-Ms}
\end{equation}%
The entropy $S_{\mathcal{M}_{s}}$ in (\ref{ent-Ms}) is the asymptotic limit
of the natural logarithm of the statistical weight $\left\langle \Delta V_{%
\mathcal{M}_{s}}\right\rangle _{\tau }$ defined on $\mathcal{M}_{s}$. Its
linear growth in time is reminiscent of the aforementioned quantum chaos
criterion. Indeed, equation (\ref{ent-Ms}) may be considered the
information-geometric analog of the Zurek-Paz chaos criterion.

In conclusion, we have shown,%
\begin{equation}
R_{\mathcal{M}_{s}}=-3N\text{, }S_{\mathcal{M}_{s}}\overset{\tau \rightarrow
\infty }{\approx }3N\lambda \tau \text{, }J_{\mathcal{M}_{S}}\overset{\tau
\rightarrow \infty }{\approx }3Ne^{\lambda \tau }\text{.}
\end{equation}%
The Ricci scalar curvature $R_{\mathcal{M}_{s}}$, the information-geometric
entropy $S_{\mathcal{M}_{s}}$ and the Jacobi vector field intensity $J_{%
\mathcal{M}_{S}}$ are proportional to the number of Gaussian-distributed
microstates of the system. This proportionality leads to the conclusion that
there exists a substantial link among these information-geometric measures
of chaoticity, namely%
\begin{equation}
R_{\mathcal{M}_{s}}\sim S_{\mathcal{M}_{s}}\sim J_{\mathcal{M}_{S}}\text{.}
\label{cool-relation}
\end{equation}%
Equation (\ref{cool-relation}), together with the information-geometric
analog of the Zurek-Paz quantum chaos criterion, equation (\ref{ent-Ms}),
represent the fundamental results of this work. We believe our theoretical
modelling scheme may be used to describe actual systems where transitions
from quantum to classical chaos scenario occur, but this requires additional
analysis. In the following section, we briefly consider some similarities
among the von Neumann, Kolmogorov-Sinai and Information-Geometrodynamical
entropies.

\section{On the von Neumann, Kolmogorov-Sinai and information
geometrodynamical Entropies}

In conventional approaches to chaos, the notion of entropy is introduced, in
both classical and quantum physics, as the missing information about the
systems fine-grained state \cite{caves, jaynes}. For a classical system,
suppose that the phase space is partitioned into very fine-grained cells of
uniform volume $\Delta v$, labelled by an index $j$. If one does not know
which cell the system occupies, one assigns probabilities $p_{j}$ to the
various cells; equivalently, in the limit of infinitesimal cells, one can
use a phase-space density $\rho \left( X_{j}\right) =\frac{p_{j}}{\Delta v}$%
. Then, in a classical chaotic evolution, the asymptotic expression of the
information needed to characterize a particular coarse-grained trajectory
out to time $\tau $ is given by the Shannon information entropy (measured in
bits),%
\begin{equation}
S_{\text{classical}}^{\left( \text{chaotic}\right) }=-\int dX\rho \left(
X\right) \log _{2}\left( \rho \left( X\right) \Delta v\right)
=-\sum_{j}p_{j}\log _{2}p_{j}\sim \mathcal{K}\tau \text{.}
\label{chao-classEnt}
\end{equation}%
where $\rho \left( X\right) $ is the phase-space density and $p_{j}=\frac{%
v_{j}}{\Delta v}$ is the probability for the corresponding coarse-grained
trajectory. $S_{\text{classical}}^{\left( \text{chaotic}\right) }$ is the
missing information about which fine-grained cell the system occupies. The
quantity $\mathcal{K}$ represents the linear rate of information increase
and it is called the Kolmogorov-Sinai entropy (or metric entropy) ($\mathcal{%
K}$ is the sum of positive Lyapunov exponents, $\mathcal{K}=\sum_{j}\lambda
_{j}$ ). $\mathcal{K}$ quantifies the degree of classical chaos. It is
worthwhile emphasizing that the quantity that grows asymptotically as $%
\mathcal{K}\tau $ is really the average of the information on the left side
of equation (\ref{chao-classEnt}). This distinction can be ignored however,
if we assume that the chaotic system has roughly constant Lyapunov exponents
over the accessible region of phase space. In quantum mechanics the
fine-grained alternatives are normalized state vectors in Hilbert space.
From a set of probabilities for various state vectors, one can construct a
density operator 
\begin{equation}
\widehat{\rho }=\sum_{j}\lambda _{j}\left\vert \psi _{j}\right\rangle
\left\langle \psi _{j}\right\vert \text{, }\widehat{\rho }\left\vert \psi
_{j}\right\rangle =\lambda _{j}\left\vert \psi _{j}\right\rangle \text{.}
\end{equation}%
The normalization of the density operator, $tr\left( \widehat{\rho }\right)
=1$, implies that the eigenvalues make up a normalized probability
distribution. The von Neumann entropy of the density operator $\widehat{\rho 
}$ (measured in bits) \cite{stenholm},%
\begin{equation}
S_{\text{quantum}}^{\left( \text{chaotic}\right) }=-tr\left( \widehat{\rho }%
\log _{2}\widehat{\rho }\right) =-\sum_{j}\lambda _{j}\log _{2}\lambda _{j}
\end{equation}%
can be thought of as the missing information about which eigenvector the
system is in. Entropy quantifies the degree of unpredictability about the
system's fine-grained state.

Recall that decoherence is the loss of phase coherence between the set of
preferred quantum states in the Hilbert space of the system due to the
interaction with the environment. Moreover, decoherence induces transitions
from quantum to classical systems. Therefore, classicality is an emergent
property of an open quantum system. Motivated by such considerations, Zurek
and Paz investigated implications of the process of decoherence for quantum
chaos.

They considered a chaotic system, a single unstable harmonic oscillator
characterized by a potential $V\left( x\right) =-\frac{\lambda x^{2}}{2}$ ($%
\lambda $ is the Lyapunov exponent), coupled to an external environment. In
the \textit{reversible classical limit }\cite{zurek2}, the von Neumann
entropy of such a system increases linearly at a rate determined by the
Lyapunov exponent,%
\begin{equation}
S_{\text{quantum}}^{\left( \text{chaotic}\right) }\left( \text{Zurek-Paz}%
\right) \overset{\tau \rightarrow \infty }{\sim }\lambda \tau \text{.}
\end{equation}%
Notice that the consideration of $3N$ uncoupled identical unstable harmonic
oscillators characterized by potentials $V_{i}\left( x\right) =-\frac{%
\lambda _{i}x^{2}}{2}$ $\left( \lambda _{i}=\lambda _{j}\text{; }i\text{, }%
j=1\text{, }2\text{,..., }3N\right) $ would simply lead to%
\begin{equation}
S_{\text{quantum}}^{\left( \text{chaotic}\right) }\left( \text{Zurek-Paz}%
\right) \overset{\tau \rightarrow \infty }{\sim }3N\lambda \tau \text{.}
\label{other-ent}
\end{equation}%
The resemblance of equations (\ref{ent-Ms}) and (\ref{other-ent}) is
remarkable. In what follows, we apply our information geometrical method to
an $n$-set (for $n=2$) of uncoupled inverted harmonic oscillators, each with
different frequency, and show we obtain asymptotic linear IGE\ growth. The
case for arbitrary $n$-set in three dimensions is presented in the Appendix.

\section{The information geometry of a $2$-\textbf{set} of inverted harmonic
oscillators (IHO)}

In this section, our objective is to characterize chaotic properties of a $2$%
-set of one-dimensional inverted harmonic oscillators, each with different
frequency $\omega _{1}\neq \omega _{2}$ using the formalism presented in
this paper. We will study the asymptotic behavior of the geometrodynamical
entropy and the functional dependence of the Ricci scalar curvature of the $%
2 $-dimensional manifold $\mathcal{M}_{IHO}^{\left( 2\right) }$ underlying
the ED model of the IHOs on the frequencies $\omega _{i\text{ }}$, $i=1$, $2$%
. Recent investigation explore the possibility of using well established
principles of inference to derive Newtonian dynamics from relevant prior
information codified into an appropriate statistical manifold \cite{cafaro5}%
. In that work the basic assumption is that there is an irreducible
uncertainty in the location of particles so that the state of a particle is
defined by a probability distribution. The corresponding configuration space
is a statistical manifold the geometry of which is defined by the
information metric. The trajectory follows from a principle of inference,
the method of Maximum Entropy. There is no need for additional "physical"
postulates such as an action principle or equation of motion, nor for the
concept of mass, momentum and of phase space, not even the notion of time.
The resulting "entropic" dynamics reproduces Newton's mechanics for any
number of particles interacting among themselves and with external fields.
Both the mass of the particles and their interactions are explained as a
consequence of the underlying statistical manifold.

In what follows, we introduce the basics of the general formalism for an $n$%
-set of IHOs. This approach is\ similar (mathematically but not
conceptually) to the geometrization of Newtonian dynamics used in the
Riemannian geometrodynamical to chaos \cite{casetti, biesiada3}.

\subsection{Informational geometrization of Newtonian dynamics}

Newtonian dynamics can be recast in the language of Riemannian geometry
applied to probability theory, namely, Information Geometry. In our case,
the system under investigation has $n$ degrees of freedom and a point on the 
$n$ dimensional configuration space manifold $\mathcal{M}_{IHO}^{\left(
n\right) }$ is parametrized by the $n$ Lagrangian coordinates $\left( \theta
_{1}\text{,...., }\theta _{n}\right) $. Moreover, the system under
investigation is described by the Lagrangian $\mathcal{L}$,%
\begin{equation}
\mathcal{L}=T\left( \overset{\cdot }{\theta }_{1}\text{,..,}\overset{\cdot }{%
\theta }_{n}\right) -\Phi \left( \theta _{1}\text{,.., }\theta _{n}\right) =%
\frac{1}{2}\delta _{ij}\overset{\cdot }{\theta }_{i}\overset{\cdot }{\theta }%
_{j}+\frac{1}{2}\overset{n}{\underset{j=1}{\sum }}\omega _{j}^{2}\theta
_{j}^{2}
\end{equation}%
so that the Hamiltonian function $\mathcal{H=}T+\Phi \equiv E$ is a constant
of motion. For the sake of simplicity, let us set $E=1$. According to the
principle of stationary action - in the form of Maupertuis - among all the
possible isoenergetic paths $\gamma \left( t\right) $ with fixed end points,
the paths that make vanish the first variation of the action functional%
\begin{equation}
\mathcal{I}=\int_{\gamma \left( t\right) }\frac{\partial \mathcal{L}}{%
\partial \overset{\cdot }{\theta }_{i}}\overset{\cdot }{\theta }_{i}d\tau
\end{equation}%
are natural motions. As the kinetic energy $T$ is a homogeneous function of
degree two, we have $2T=\overset{\cdot }{\theta }_{i}\frac{\partial \mathcal{%
L}}{\partial \overset{\cdot }{\theta }_{i}}$, and Maupertuis' principle reads%
\begin{equation}
\delta \mathcal{I}=\delta \int_{\gamma \left( t\right) }2Td\tau =0\text{.}
\end{equation}%
The manifold $\mathcal{M}_{IHO}^{\left( n\right) }$ is naturally given a
proper Riemannian structure. In fact, let us consider the matrix%
\begin{equation}
g_{ij}\left( \theta _{1}\text{,.., }\theta _{n}\right) =\left[ 1-\Phi \left(
\theta _{1}\text{,.., }\theta _{n}\right) \right] \delta _{ij}
\label{iho-metric2}
\end{equation}%
so that Maupertuis' principle becomes%
\begin{eqnarray}
\delta \int_{\gamma \left( t\right) }Tdt &=&\delta \int_{\gamma \left(
t\right) }\left( T^{2}\right) ^{\frac{1}{2}}d\tau =\delta \int_{\gamma
\left( t\right) }\left\{ \left[ 1-\Phi \left( \theta _{1}\text{,..,}\theta
_{n}\right) \right] \delta _{ij}\overset{\cdot }{\theta }_{i}\overset{\cdot }%
{\theta }_{j}\right\} ^{\frac{1}{2}}  \notag \\
&=&\delta \int_{\gamma \left( t\right) }\left( g_{ij}\overset{\cdot }{\theta 
}_{i}\overset{\cdot }{\theta }_{j}\right) ^{\frac{1}{2}}d\tau =\delta
\int_{\gamma \left( s\right) }ds=0\text{, }ds^{2}=g_{ij}d\theta ^{i}d\theta
^{j}
\end{eqnarray}%
thus natural motions are geodesics of $\mathcal{M}_{IHO}^{\left( n\right) }$%
, provided we define $ds$ as its arclength. The metric tensor $g_{J}\left(
\cdot \text{, }\cdot \right) $ of $\mathcal{M}_{IHO}^{\left( n\right) }$ is
then defined by%
\begin{equation}
g=g_{ij}d\theta ^{i}\otimes d\theta ^{j}  \label{iho-metric}
\end{equation}%
where $\left( d\theta ^{1}\text{,....., }d\theta ^{n}\right) $ is a natural
base of $T_{\theta }^{\ast }\mathcal{M}_{IHO}^{\left( n\right) }$ - the
cotangent space at the point $\theta $ - in the local chart $\left( \theta
^{1}\text{,...., }\theta ^{n}\right) $. This is known as the Jacobi metric
(or kinetic energy metric). Denoting by $\nabla $ the canonical Levi-Civita
connection, the geodesic equation%
\begin{equation}
\nabla _{\overset{\cdot }{\gamma }}\overset{\cdot }{\gamma }=0
\end{equation}%
becomes, in the local chart $\left( \theta ^{1}\text{,...., }\theta
^{n}\right) $,%
\begin{equation}
\frac{d^{2}\theta ^{i}}{ds^{2}}+\Gamma _{jk}^{i}\frac{d\theta ^{j}}{ds}\frac{%
d\theta ^{k}}{ds}=0
\end{equation}%
where the Christoffel coefficients are the components of $\nabla $ defined by%
\begin{equation}
\Gamma _{jk}^{i}=\left\langle d\theta ^{i}\text{, }\nabla
_{j}e_{k}\right\rangle =\frac{1}{2}g^{im}\left( \partial _{j}g_{km}+\partial
_{k}g_{mj}-\partial _{m}g_{jk}\right) \text{,}  \label{iho-connection}
\end{equation}%
with $\partial _{i}=\frac{\partial }{\partial \theta ^{i}}$. Since $%
g_{ij}\left( \theta _{1}\text{,.., }\theta _{n}\right) =\left[ 1-\Phi \left(
\theta _{1}\text{,.., }\theta _{n}\right) \right] \delta _{ij}$, from the
geodesic equation we obtain%
\begin{equation}
\frac{d^{2}\theta ^{i}}{ds^{2}}+\frac{1}{2\left( 1-\Phi \right) }\left[ 2%
\frac{\partial \left( 1-\Phi \right) }{\partial \theta _{j}}\frac{d\theta
^{j}}{ds}\frac{d\theta ^{i}}{ds}-g^{ij}\frac{\partial \left( 1-\Phi \right) 
}{\partial \theta _{j}}g_{km}\frac{d\theta ^{k}}{ds}\frac{d\theta ^{m}}{ds}%
\right] =0\text{,}  \label{inter}
\end{equation}%
whereupon using $ds^{2}=\left( 1-\Phi \right) ^{2}d\tau ^{2}$, we verify
that (\ref{inter}) reduces to%
\begin{equation}
\frac{d^{2}\theta ^{i}}{d\tau ^{2}}+\frac{\partial \Phi \left( \theta _{1}%
\text{,.., }\theta _{n}\right) }{\partial \theta _{i}}=0\text{, }i=1\text{%
,..., }n\text{.}  \label{Newton}
\end{equation}%
Equation (\ref{Newton}) are Newton's equations. It is worthwhile emphasizing
that the transformation to geodesic motion on a curved statistical manifold
is obtained in two key steps: the \textit{conformal transformation of the
metric}, $\delta _{ij}\rightarrow g_{ij}=\left( 1-\Phi \right) $ $\delta
_{ij}$ and, the \textit{rescaling of the temporal evolution parameter}, $%
d\tau ^{2}\rightarrow ds^{2}=2\left( 1-\Phi \right) ^{2}d\tau ^{2}$.

\subsection{The $2$-\textbf{set} of inverted anisotropic one-dimensional
harmonic oscillators}

As a simple physical example, we examine the IG associated with a $2$-set
configuration of IHOs. In this case, the metric tensor $g_{ij}$ appearing in
(\ref{iho-metric2}) takes the form%
\begin{equation}
g_{ij}\left( \theta _{1}\text{, }\theta _{2}\right) =\left[ 1-\Phi \left(
\theta _{1}\text{, }\theta _{2}\right) \right] \cdot \delta _{ij}\left(
\theta _{1}\text{, }\theta _{2}\right) \text{ with }i\text{, }j=1\text{, }2%
\text{.}
\end{equation}%
where the function $\Phi \left( \theta _{1}\text{, }\theta _{2}\right) $ is
given by,%
\begin{equation}
\Phi \left( \theta _{1}\text{, }\theta _{2}\right) =\overset{2}{\underset{j=1%
}{\sum }}\Phi _{j}\left( \theta _{j}\right) \text{, }\Phi _{j}\left( \theta
_{j}\right) =-\frac{1}{2}\omega _{j}^{2}\theta _{j}^{2}\text{.}
\end{equation}%
Hence the metric tensor $g_{ij}$ on $\mathcal{M}_{IHO}^{\left( 2\right) }$
becomes,%
\begin{equation}
g_{ij}=\left( 
\begin{array}{cc}
1+\frac{1}{2}\left( \omega _{1}^{2}\theta _{1}^{2}+\omega _{2}^{2}\theta
_{2}^{2}\right) & 0 \\ 
0 & 1+\frac{1}{2}\left( \omega _{1}^{2}\theta _{1}^{2}+\omega _{2}^{2}\theta
_{2}^{2}\right)%
\end{array}%
\right) \text{.}
\end{equation}%
Using the standard definition of the Ricci scalar (\ref{ricci-scalar}), we
obtain%
\begin{equation}
R_{\mathcal{M}_{IHO}^{\left( 2\right) }}\left( \omega _{1}\text{, }\omega
_{2}\right) =\frac{4\left( \theta _{1}^{2}\omega _{1}^{4}+\theta
_{2}^{2}\omega _{2}^{4}\right) -4\left( \theta _{1}^{2}+\theta
_{2}^{2}\right) \omega _{1}^{2}\omega _{2}^{2}-8\left( \omega
_{1}^{2}+\omega _{2}^{2}\right) }{\left( \theta _{1}^{2}\omega
_{1}^{2}+\theta _{2}^{2}\omega _{2}^{2}+2\right) ^{3}}\text{.}
\label{ricciS-iho}
\end{equation}%
In the limit of a flat frequency spectrum, $\omega _{1}=\omega _{2}=\omega $%
, the scalar curvature (\ref{ricciS-iho}) is constantly negative, 
\begin{equation}
R_{\mathcal{M}_{IHO}^{\left( 2\right) }}\left( \omega \right) =\frac{%
-16\omega ^{2}}{\left[ 2+\left( \theta _{1}^{2}+\theta _{2}^{2}\right)
\omega ^{2}\right] ^{3}}<0\text{, }\forall \omega \geq 0\text{.}
\end{equation}%
However, in presence of distinct frequency values, $\omega _{1}\neq \omega
_{2}$, it is possible to properly choose the $\omega $'s so that $R_{%
\mathcal{M}_{IHO}^{\left( 2\right) }}\left( \omega _{1}\text{, }\omega
_{2}\right) $ becomes either negative or positive. In addition, we notice
that the manifold underlying the IHO model is anisotropic since its
associated Weyl projective curvature tensor components are non-vanishing.
For the special case, $\omega _{1}=\omega _{2}$, we obtain%
\begin{equation}
W_{1212}\left( \omega \right) =\frac{8\omega ^{4}\left( \theta
_{1}^{2}+\theta _{2}^{2}\right) +2\omega ^{6}\left( \theta _{1}^{4}+\theta
_{2}^{4}\right) +4\omega ^{6}\theta _{1}^{2}\theta _{2}^{2}}{\left( \theta
_{1}^{2}\omega ^{2}+\theta _{2}^{2}\omega ^{2}+2\right) ^{3}}\text{.}
\end{equation}%
Clearly, the frequency parameter $\omega $ drives the degree of anisotropy
of the statistical manifold $\mathcal{M}_{IHO}^{\left( 2\right) }$ and, as
expected, in the limit of vanishing $\omega $, we recover the flat ($R=0$),
isotropic ($W=0$) Euclidean manifold characterized by metric $\delta _{ij}$.
This result is a concrete example of the fact that conformal transformations
change the degree of anisotropy of the ambient statistical manifold
underlying the Newtonian dynamics. Our only remaining task is to compute the
information geometrodynamical entropy $S_{\mathcal{M}_{IHO}^{\left( 2\right)
}}\left( \tau \text{; }\omega _{1}\text{, }\omega _{2}\right) $, defined as%
\begin{equation}
S_{\mathcal{M}_{IHO}^{\left( 2\right) }}\left( \tau \text{; }\omega _{1}%
\text{, }\omega _{2}\right) \overset{\text{def}}{=}\underset{\tau
\rightarrow \infty }{\lim }\log \left[ \left\langle \Delta V_{\mathcal{M}%
_{IHO}^{\left( 2\right) }}\left( \tau \text{; }\omega _{1}\text{, }\omega
_{2}\right) \right\rangle _{\tau }\right] \text{.}  \label{iho-entropy}
\end{equation}%
The quantity $\left\langle \Delta V_{\mathcal{M}_{IHO}^{\left( 2\right)
}}\left( \tau \text{; }\omega _{1}\text{, }\omega _{2}\right) \right\rangle
_{\tau }$ appearing in (\ref{iho-entropy}) is the average volume element,
defined by%
\begin{equation}
\left\langle \Delta V_{\mathcal{M}_{IHO}^{\left( 2\right) }}\left( \tau 
\text{; }\omega _{1}\text{, }\omega _{2}\right) \right\rangle _{\tau }=\frac{%
1}{\tau }\dint\limits_{0}^{\tau }\Delta V_{\mathcal{M}_{IHO}^{\left(
2\right) }}\left( \tau \text{; }\omega _{1}\text{, }\omega _{2}\right) d\tau
^{\prime }\text{,}
\end{equation}%
with the statistical volume element $\Delta V_{\mathcal{M}_{IHO}^{\left(
2\right) }}$ given by%
\begin{eqnarray}
\Delta V_{\mathcal{M}_{IHO}^{\left( 2\right) }}\left( \tau \text{; }\omega
_{1}\text{, }\omega _{2}\right) &=&\underset{\left\{ \vec{\theta}^{\prime
}\right\} }{\int }\left[ 1+\frac{1}{2}\left( \omega _{1}^{2}\theta
_{1}^{\prime 2}+\omega _{2}^{2}\theta _{2}^{\prime 2}\right) \right] d\theta
_{1}^{\prime }d\theta _{2}^{\prime }  \label{delV} \\
&&  \notag \\
&&\overset{\tau \rightarrow \infty }{\approx }\frac{1}{6}\theta _{1}^{\prime
}\theta _{2}^{\prime }\left( \omega _{1}^{2}\theta _{1}^{\prime 2}+\omega
_{2}^{2}\theta _{2}^{\prime 2}\right) \text{.}  \notag
\end{eqnarray}%
Recall that the two Newtonian equations of motion for each inverted harmonic
oscillator are given by,%
\begin{equation}
\frac{d^{2}\theta _{j}}{d\tau ^{2}}-\omega _{j}^{2}\theta _{j}=0\text{, }%
\forall j=1\text{, }2\text{.}
\end{equation}%
Hence, the asymptotic behavior of such macrovariables on manifold $\mathcal{M%
}_{IHO}^{\left( 2\right) }$ is given by,%
\begin{equation}
\theta _{j}\left( \tau \right) \overset{\tau \rightarrow \infty }{\approx }%
\Xi _{j}e^{\omega _{j}\tau }\text{, }\Xi _{j}\in 
%TCIMACRO{\U{211d} }%
%BeginExpansion
\mathbb{R}
%EndExpansion
\text{, }\forall j=1\text{, }2\text{.}
\end{equation}%
Substituting $\theta _{1}\left( \tau ^{\prime }\right) =\Xi _{1}e^{\omega
_{1}\tau ^{\prime }}$ and $\theta _{2}\left( \tau ^{\prime }\right) =\Xi
_{2}e^{\omega _{2}\tau ^{\prime }}$ into (\ref{delV}), we obtain%
\begin{equation}
\Delta V_{\mathcal{M}_{IHO}^{\left( 2\right) }}\left( \tau \text{; }\omega
_{1}\text{, }\omega _{2}\right) \overset{\tau \rightarrow \infty \text{ }}{%
\approx }\frac{\Xi _{1}\Xi _{2}}{6}e^{\left( \omega _{1}+\omega _{2}\right)
\tau }\left( \Xi _{1}^{2}e^{2\omega _{1}\tau }\omega _{1}^{2}+\Xi
_{2}^{2}e^{2\omega _{2}\tau }\omega _{2}^{2}\right) \text{.}  \label{above}
\end{equation}%
By direct computation, we find the average of (\ref{above}) is given by,%
\begin{equation}
\left\langle \Delta V_{\mathcal{M}_{IHO}^{\left( 2\right) }}\left( \tau 
\text{; }\omega _{1}\text{, }\omega _{2}\right) \right\rangle _{\tau }%
\overset{\tau \rightarrow \infty \text{ }}{\approx }\frac{1}{\tau }%
\dint\limits_{0}^{\tau }\left[ \frac{\Xi _{1}\Xi _{2}}{6}e^{\left( \omega
_{1}+\omega _{2}\right) \tau ^{\prime }}\left( \Xi _{1}^{2}e^{2\omega
_{1}\tau ^{\prime }}\omega _{1}^{2}+\Xi _{2}^{2}e^{2\omega _{2}\tau ^{\prime
}}\omega _{2}^{2}\right) \right] d\tau ^{\prime }\text{.}
\end{equation}%
Assuming as a working hypothesis that $\Xi _{1}=\Xi _{2}=\Xi $, we obtain%
\begin{equation}
\frac{1}{\tau }\dint\limits_{0}^{\tau }\left[ \frac{\Xi _{1}\Xi _{2}}{6}%
e^{\left( \omega _{1}+\omega _{2}\right) \tau ^{\prime }}\left( \Xi
_{1}^{2}e^{2\omega _{1}\tau ^{\prime }}\omega _{1}^{2}+\Xi
_{2}^{2}e^{2\omega _{2}\tau ^{\prime }}\omega _{2}^{2}\right) \right] d\tau
^{\prime }=\left\{ 
\begin{array}{c}
\frac{1}{12}\Xi ^{6}\omega \frac{\exp \left( 4\omega \tau \right) }{\tau }%
\text{, if }\omega _{1}=\omega _{2}\text{,} \\ 
\\ 
\frac{1}{18}\Xi ^{6}\omega _{1}\frac{\exp \left( 3\omega _{1}\tau \right) }{%
\tau }\text{, if }\omega _{1}\gg \omega _{2}\text{,} \\ 
\\ 
\frac{1}{18}\Xi ^{6}\omega _{2}\frac{\exp \left( 3\omega _{2}\tau \right) }{%
\tau }\text{, if }\omega _{2}\gg \omega _{1}\text{.}%
\end{array}%
\right. \text{.}  \label{this}
\end{equation}%
Finally, substituting (\ref{this}) in (\ref{iho-entropy}), we obtain%
\begin{equation}
S_{\mathcal{M}_{IHO}^{\left( 2\right) }}\left( \tau \text{; }\omega _{1}%
\text{, }\omega _{2}\right) \overset{\tau \rightarrow \infty }{\propto }%
\left\{ 
\begin{array}{c}
2\omega \tau \text{, if }\omega _{1}=\omega _{2}\text{,} \\ 
\\ 
\omega _{1}\tau \text{, if }\omega _{1}\gg \omega _{2}\text{,} \\ 
\\ 
\omega _{2}\tau \text{, if }\omega _{2}\gg \omega _{1}\text{.}%
\end{array}%
\right. \text{.}  \label{iho-entropy1}
\end{equation}%
It is clear that the information-geometrodynamical entropy $S_{\mathcal{M}%
_{IHO}^{\left( 2\right) }}\left( \tau \text{; }\omega _{1}\text{, }\omega
_{2}\right) $ exhibits classical linear behavior in the asymptotic limit,
with proportionality coefficient $\Omega =$ $\omega _{1}+\omega _{2}$,%
\begin{equation}
S_{\mathcal{M}_{IHO}^{\left( 2\right) }}\left( \tau \text{; }\omega _{1}%
\text{, }\omega _{2}\right) \overset{\tau \rightarrow \infty }{\propto }%
\Omega \tau \text{.}  \label{sopra}
\end{equation}%
Equation (\ref{sopra}) expresses the asymptotic linear growth of our
information geometrodynamical entropy for the IHO system considered. This
result (for $n=2$) extends the result of Zurek-Paz (\ref{other-ent}). This
result, together with the authors previous works \cite{cafaro2, cafaro3}
lend substantial support for the IG approach advocated in the present
article.

\section{Final remarks}

A Gaussian ED statistical model has been constructed on a $6N$-dimensional
statistical manifold $\mathcal{M}_{s}$. The macro-coordinates on the
manifold are represented by the expectation values of microvariables
associated with Gaussian distributions. The geometric structure of $\mathcal{%
M}_{s}$ was studied in detail. It was shown that $\mathcal{M}_{s}$ is a
curved manifold of constant negative Ricci curvature $-3N$ . The geodesics
of the ED model are hyperbolic curves on $\mathcal{M}_{s}$. A study of the
stability of geodesics on $\mathcal{M}_{s}$ was presented. The notion of
statistical volume elements was introduced to investigate the asymptotic
behavior of a one-parameter family of neighboring volumes $\mathcal{F}_{V_{%
\mathcal{M}_{s}}}\left( \lambda \right) \equiv \left\{ V_{\mathcal{M}%
_{s}}\left( \tau \text{; }\lambda \right) \right\} _{\lambda \in 
%TCIMACRO{\U{211d} }%
%BeginExpansion
\mathbb{R}
%EndExpansion
^{+}}$. An information-geometric analog of the Zurek-Paz chaos criterion was
suggested. It was shown that the behavior of geodesics is characterized by
exponential instability that leads to chaotic scenarios on the curved
statistical manifold. These conclusions are supported by a study based on
the geodesic deviation equations and on the asymptotic behavior of the
Jacobi vector field intensity $J_{\mathcal{M}_{s}}$ on $\mathcal{M}_{s}$. A
Lyapunov exponent analog similar to that appearing in the Riemannian
geometric approach to chaos was suggested as an indicator of chaoticity. On
the basis of our analysis a relationship among an entropy-like quantity,
chaoticity and curvature is proposed, suggesting to interpret the
statistical curvature as a measure of the entropic dynamical chaoticity.

The results obtained in this work are significant, in our opinion, since a
rigorous relation among curvature, Lyapunov exponents and Kolmogorov-Sinai
entropy is still under investigation \cite{kawabe}. In addition, there does
not exist a well defined unifying characterization of chaos in classical and
quantum physics \cite{caves} due to fundamental differences between the two
theories. In addition, the role of curvature in statistical inference is
even less understood. The meaning of statistical curvature for a
one-parameter model in inference theory was introduced in \cite{efron}.
Curvature served as an important tool in the asymptotic theory of
statistical estimation. Therefore the implications of this work is twofold.
Firstly, it helps understanding possible future use of the statistical
curvature in modelling real processes by relating it to conventionally
accepted quantities such as entropy and chaos. On the other hand, it serves
to cast what is already known in physics regarding curvature in a new light
as a consequence of its proposed link with inference.

As a simple physical example, we considered the information-geometry $%
\mathcal{M}_{IHO}^{\left( 2\right) }$ associated with a $2$-set
configuration of inverted harmonic oscillators. It was determined that in
the limit of a flat frequency spectrum ($\omega _{1}=\omega _{2}=\omega $),
the scalar curvature $R_{\mathcal{M}_{IHO}^{\left( 2\right) }}\left( \omega
_{1}\text{, }\omega _{2}\right) $ is constantly negative. In the case of
distinct frequencies, i.e., $\omega _{1}\neq \omega _{2}$, it is possible -
for appropriate choices of $\omega _{1}$ and $\omega _{2}$ - to obtain
either negative or positive values of $R_{\mathcal{M}_{IHO}^{\left( 2\right)
}}\left( \omega _{1}\text{, }\omega _{2}\right) $. Moreover, it was shown
that $\mathcal{M}_{IHO}^{\left( 2\right) }$ is an anisotropic manifold since
the Weyl projective curvature tensor has a non-vanishing component $W_{1212}$%
. It was found that the information geometrodynamical entropy of the IHO
system exhibits asymptotic linear growth. This IHO example is generalized to
arbitrary values of $n$ in the Appendix.

The descriptions of a classical chaotic system of arbitrary interacting
degrees of freedom, deviations from Gaussianity and chaoticity arising from
fluctuations of positively curved statistical manifolds are being
investigated \cite{cafaro4}. The work here presented is shown to be useful
to investigate chaotic quantum spectra arising, for instance, from the
Poisson and Wigner-Dyson quantum level spacing distributions \cite%
{wigner-dyson, berry, cafaro6}. We remark that based on the results obtained
from the chosen ED models, it is not unreasonable to think that should the
correct variables describing the true degrees of freedom of a physical
system be identified, perhaps deeper insights into the foundations of models
of physics and reasoning (and their relationship to each other) may be
uncovered.

\section{Acknowledgement}

The authors are grateful to Prof. Ariel Caticha and Adom Giffin for useful
comments.

\section{Appendix}

\subsection{The $n$-\textbf{set} of inverted anisotropic three-dimensional
harmonic oscillators}

We now generalize the results obtained in this article for the $n$-set of
IHOs. The information metric on the $3n$-dimensional statistical manifold $%
\mathcal{M}_{IHO}^{\left( 3n\right) }$ is given by%
\begin{equation}
g_{ij}\left( \theta _{1}\text{,...., }\theta _{3n}\right) =\left[ 1-\Phi
\left( \theta _{1}\text{,...., }\theta _{3n}\right) \right] \cdot \delta
_{ij}\left( \theta _{1}\text{,...., }\theta _{3n}\right) \text{,}
\end{equation}%
where%
\begin{equation}
\Phi \left( \theta _{1}\text{,...., }\theta _{3n}\right) =\overset{3n}{%
\underset{j=1}{\sum }}\Phi _{j}\left( \theta _{j}\right) \text{, }\Phi
_{j}\left( \theta _{j}\right) =-\frac{1}{2}\omega _{j}^{2}\theta _{j}^{2}%
\text{.}
\end{equation}%
The information geometrodynamical entropy $S_{\mathcal{M}_{IHO}^{\left(
3n\right) }}\left( \tau \text{; }\omega _{1}\text{,.., }\omega _{3n}\right) $
is defined as%
\begin{equation}
S_{\mathcal{M}_{IHO}^{\left( 3n\right) }}\left( \tau \text{; }\omega _{1}%
\text{,.., }\omega _{3n}\right) \overset{\text{def}}{=}\underset{\tau
\rightarrow \infty }{\lim }\log \left[ \left\langle \Delta V_{\mathcal{M}%
_{IHO}^{\left( 3n\right) }}\left( \tau \text{; }\omega _{1}\text{,.., }%
\omega _{3n}\right) \right\rangle _{\tau }\right] \text{,}  \label{gen-ent}
\end{equation}%
where the average volume element $\Delta V_{\mathcal{M}_{IHO}^{\left(
3n\right) }}$ is given by%
\begin{equation}
\left\langle \Delta V_{\mathcal{M}_{IHO}^{\left( 3n\right) }}\left( \tau 
\text{; }\omega _{1}\text{,.., }\omega _{3n}\right) \right\rangle _{\tau }=%
\frac{1}{\tau }\dint\limits_{0}^{\tau }\Delta V_{\mathcal{M}_{IHO}^{\left(
3n\right) }}\left( \tau ^{\prime }\text{; }\omega _{1}\text{,.., }\omega
_{3n}\right) d\tau ^{\prime }\text{,}  \label{inter2}
\end{equation}%
and the statistical volume element $\Delta V_{\mathcal{M}_{IHO}^{\left(
3n\right) }}$ is defined as%
\begin{equation}
\Delta V_{\mathcal{M}_{IHO}^{\left( 3n\right) }}\left( \tau ^{\prime }\text{%
; }\omega _{1}\text{,.., }\omega _{n}\right) =\underset{\left\{ \vec{\theta}%
^{\prime }\right\} }{\int }d^{3n}\vec{\theta}^{\prime }\left( 1+\frac{1}{2}%
\underset{j=1}{\overset{3n}{\sum }}\omega _{j}^{2}\theta _{j}^{\prime
2}\right) ^{\frac{3n}{2}}\text{.}  \label{inter3}
\end{equation}%
Substituting (\ref{inter2}) and (\ref{inter3}) in (\ref{gen-ent}) we obtain
the general expression for $S_{\mathcal{M}_{IHO}^{\left( 3n\right) }}\left(
\tau \text{; }\omega _{1}\text{,.., }\omega _{3n}\right) $, 
\begin{equation}
S_{\mathcal{M}_{IHO}^{\left( 3n\right) }}\left( \tau \text{; }\omega _{1}%
\text{,.., }\omega _{3n}\right) \overset{\text{def}}{=}\underset{\tau
\rightarrow \infty }{\lim }\log \left\{ \frac{1}{\tau }\int_{0}^{\tau }\left[
\underset{\left\{ \vec{\theta}^{\prime }\right\} }{\int }d^{3n}\vec{\theta}%
^{\prime }\left( 1+\frac{1}{2}\underset{j=1}{\overset{3n}{\sum }}\omega
_{j}^{2}\theta _{j}^{\prime 2}\right) ^{\frac{3n}{2}}\right] d\tau ^{\prime
}\right\} \text{.}  \label{inter4}
\end{equation}%
To evaluate (\ref{inter4}) we observe $\Delta V_{\mathcal{M}_{IHO}^{\left(
3n\right) }}$ can be written as%
\begin{eqnarray}
\Delta V_{\mathcal{M}_{IHO}^{\left( 3n\right) }}\left( \tau ^{\prime }\text{%
; }\omega _{1}\text{,.., }\omega _{3n}\right) &=&\underset{\left\{ \vec{%
\theta}^{\prime }\right\} }{\int }d^{3n}\vec{\theta}^{\prime }\left( 1+\frac{%
1}{2}\underset{j=1}{\overset{3n}{\sum }}\omega _{j}^{2}\theta _{j}^{\prime
2}\right) ^{\frac{3n}{2}}\text{,}  \notag \\
&&  \notag \\
&=&\int d\theta _{1}^{\prime }\int d\theta _{2}^{\prime }\text{...}\int
d\theta _{3n-1}^{\prime }\left[ \int \left( 1+\frac{1}{2}\underset{j=1}{%
\overset{3n}{\sum }}\omega _{j}^{2}\theta _{j}^{\prime 2}\right) ^{\frac{3n}{%
2}}d\theta _{3n}^{\prime }\right] \text{,}  \notag \\
&&  \notag \\
&&\overset{\text{ }}{\approx }\frac{1}{3n}\frac{1}{2^{\frac{3n}{2}}}\left( 
\overset{3n}{\underset{i=1}{\Pi }}\theta _{i}^{\prime }\right) \left[ 
\underset{j=1}{\overset{3n}{\sum }}\omega _{j}^{2}\theta _{j}^{\prime 2}%
\right] ^{\frac{3n}{2}}\text{.}
\end{eqnarray}%
Since the $n$-Newtonian equations of motions for each IHO are given by%
\begin{equation}
\frac{d^{2}\theta _{j}}{d\tau ^{2}}-\omega _{j}^{2}\theta _{j}=0\text{, }%
\forall j=1\text{,..., }3n\text{,}
\end{equation}%
the asymptotic behavior of such macrovariables on manifold $\mathcal{M}%
_{IHO}^{\left( 3n\right) }$ is given by%
\begin{equation}
\theta _{j}\left( \tau \right) \overset{\tau \rightarrow \infty }{\approx }%
\Xi _{j}e^{\omega _{j}\tau }\text{, }\Xi _{j}\in 
%TCIMACRO{\U{211d} }%
%BeginExpansion
\mathbb{R}
%EndExpansion
\text{, }\forall j=1\text{,..., }3n\text{.}
\end{equation}%
We therefore obtain%
\begin{equation}
\Delta V_{\mathcal{M}_{IHO}^{\left( 3n\right) }}\left( \tau \text{; }\omega
_{1}\text{,...., }\omega _{3n}\right) \overset{\tau \rightarrow \infty \text{
}}{\approx }\frac{1}{3n}\frac{1}{2^{\frac{3n}{2}}}\left( \underset{i=1}{%
\overset{3n}{\Pi }}\Xi _{i}\right) \cdot \exp \left( \overset{3n}{\underset{%
i=1}{\sum }}\omega _{i}\tau \right) \left[ \underset{j=1}{\overset{3n}{\sum }%
}\Xi _{j}^{2}e^{2\omega _{j}\tau }\omega _{j}^{2}\right] ^{\frac{3n}{2}}%
\text{.}  \label{inter5}
\end{equation}%
Upon averaging (\ref{inter5}) we find%
\begin{equation}
\left\langle \Delta V_{\mathcal{M}_{IHO}^{\left( 3n\right) }}\left( \tau 
\text{; }\omega _{1}\text{,...., }\omega _{3n}\right) \right\rangle _{\tau }%
\overset{\tau \rightarrow \infty \text{ }}{\approx }\frac{1}{\tau }%
\dint\limits_{0}^{\tau }\left\{ \frac{1}{3n}\frac{1}{2^{\frac{3n}{2}}}\left( 
\underset{i=1}{\overset{3n}{\Pi }}\Xi _{i}\right) \cdot \exp \left( \Omega
\tau ^{\prime }\right) \left[ \underset{j=1}{\overset{3n}{\sum }}\Xi
_{j}^{2}e^{2\omega _{j}\tau ^{\prime }}\omega _{j}^{2}\right] ^{\frac{3n}{2}%
}\right\} d\tau ^{\prime }\text{.}
\end{equation}%
where $\Omega =\overset{3n}{\underset{i=1}{\sum }}\omega _{i}$. As a working
hypothesis, we assume $\Xi _{i}=\Xi _{j}\equiv \Xi $ $\forall i$, $j=1$,.., $%
3n$. Furthermore, assume that $n\rightarrow \infty $ so that the spectrum of
frequencies becomes continuum and, as an additional working hypothesis,
assume this spectrum is linearly distributed,%
\begin{equation}
\rho \left( \omega \right) =\omega \text{ with}\underset{0}{\overset{\Omega
_{\text{cut-off}}}{\int }}\rho \left( \omega \right) d\omega =1\text{, }%
\Omega _{\text{cut-off}}=\xi \Omega \text{, }\xi \in 
%TCIMACRO{\U{211d} }%
%BeginExpansion
\mathbb{R}
%EndExpansion
\text{ }
\end{equation}%
Therefore, we obtain%
\begin{equation}
\left\langle \Delta V_{\mathcal{M}_{IHO}^{\left( 3n\right) }}\left( \tau 
\text{; }\omega _{1}\text{,...., }\omega _{3n}\right) \right\rangle _{\tau }%
\overset{\tau \rightarrow \infty \text{ }}{\approx }\frac{1}{3n}\frac{1}{2^{%
\frac{3n}{2}}}\Xi ^{6n}\left( \frac{\xi ^{2}\Omega ^{2}}{2}\right) ^{\frac{3n%
}{2}}\frac{\exp \left( \frac{3}{2}n\xi \Omega \tau \right) }{\tau }\text{.}
\label{inter6}
\end{equation}%
Finally, substituting (\ref{inter6}) into (\ref{gen-ent}), we obtain the
remarkable result%
\begin{equation}
S_{\mathcal{M}_{IHO}^{\left( 3n\right) }}\left( \tau \text{; }\omega _{1}%
\text{,.., }\omega _{3n}\right) \overset{\tau \rightarrow \infty }{\propto }%
\Omega \tau \text{, }\Omega =\overset{3n}{\underset{i=1}{\sum }}\omega _{i}%
\text{.}  \label{Fin}
\end{equation}%
Equation (\ref{Fin}) displays the asymptotic, linear information
geometrodynamical entropy growth of the generalized $n$-set of inverted
harmonic oscillators and extends the result of Zurek-Paz to an arbitrary set
of anisotropic inverted harmonic oscillators \cite{zurek}.

\end{document}